\documentclass[pre,twocolumn,superscriptaddress,floatfix,showpacs]{revtex4}

\usepackage[latin2]{inputenc}
\usepackage{amsmath}
\usepackage{amssymb}
\usepackage{epsfig}
\usepackage{color}
\usepackage{bm}
\usepackage{cancel}
\begin{document}

\title{Long-range interactions in randomly driven granular fluids}

\author{M.\ Reza Shaebani}
\email{shaebani@lusi.uni-sb.de}
\affiliation{Department of Theoretical Physics, Saarland University, 
D-66041 Saarbruecken, Germany}
\author{Jalal Sarabadani}
\altaffiliation{present address: Max Planck Institute 
for Polymer Research, Ackermannweg 10, D-55128 Mainz, Germany}
\email{jalal.sarabadani@iasbs.ac.ir}
\affiliation{Department of Physics, Institute for Advanced 
Studies in Basic Sciences (IASBS), Zanjan 45137-66731, Iran} 
\author{Dietrich E.\ Wolf}
\affiliation{Department of Theoretical Physics, University of
Duisburg-Essen, 47048 Duisburg, Germany}

\date{\today}

\begin{abstract}
We study the long-range spatial correlations in the nonequilibrium 
steady state of a randomly driven granular fluid with the emphasis 
on obtaining the explicit form of the static structure factors. The 
presence of immobile particles immersed in such a fluidized bed of 
fine particles leads to the confinement of the fluctuation spectrum 
of the hydrodynamic fields, which results in effective long-range 
interactions between the intruders. The analytical predictions are 
in agreement with the results of discrete element method simulations. 
By changing the shape and orientation of the intruders, we 
address how the effective force is affected by small changes in 
the boundary conditions.
\end{abstract}

\pacs{45.70.Mg, 05.40.-a}

\maketitle

\section{Introduction}
\label{Introduction-Section}

A freely evolving gas of inelastically colliding particles behaves 
differently from a regular gas due to the dissipative nature of 
collisions. Starting with a uniformly distributed granular gas, 
the system first enters a homogeneous cooling state, where the 
kinetic energy decays as $t^{-2}$, known as Haff's law \cite{Haff83}. 
However, this regime is unstable against spatial fluctuations of 
collision frequency, leading to a crossover to a regime where the 
energy decreases as $t^{-d/2}$ in \emph{d} dimensions and eventually 
dense clusters appear \cite{Brito98,Goldhirsch93,Efrati05,Cattuto04}. 
In order to maintain the dynamics, energy must be pumped into the 
system from an external driving source e.g.\ by shaking, vibrating, 
or shearing. As a result, the energy input competes against the 
energy dissipation due to inelastic collisions, and the system is 
driven into a steady state. Since the energy insertion is spatially 
inhomogeneous in the above mentioned \emph{boundary heating} methods, 
a \emph{uniformly heating} approach had been proposed as a paradigm 
allowing to study a homogeneous stationary state 
\cite{Williams96,Peng98,vanNoije99}; The system is supposed to be in 
contact with a thermal bath which supplies energy by means of 
instantaneous random forces acting on all particles. The resulting 
dynamics, in two dimensions, is reminiscent of the motion of thin 
disks on an air table in quasi-2D experiments \cite{Oger96}. The 
uniformly driven granular fluid reaches a \emph{nonequilibrium} 
steady state due to the random nature of energy gain and loss, 
and exhibits \emph{long-range} hydrodynamic correlations. These 
spatial correlations decay with distance as $r^{-1}$ and $\ln(1/r)$ 
in three and two dimensions, respectively \cite{vanNoije99}. 
Similar long-range spatial correlations emerge in freely cooling 
\cite{vanNoije97} or sheared \cite{Otsuki09} systems of slightly 
inelastic granular fluids, even though the anisotropic couplings 
between the number density and momentum density fluctuations 
influences the behavior in the sheared case, leading to a different 
power-law asymptotic form at large $r$ \cite{Otsuki09}.

In a thermally fluctuating medium with long-range spatial 
correlations, one expects that the presence of big intruder 
particles modifies the boundary conditions and, hence, leads 
to the modification of the hydrodynamic fluctuations spectrum. 
As a result, effective interactions between intruders appear 
which are long-range, in spite of the short-range nature of 
hard-core interparticle forces. Such interactions are often 
considered as classical variants of the Casimir force 
\cite{Casimir48,Kardar99}. Recent experiments 
\cite{Aumaitre01,Zuriguel06,Villanueva10,GarciaTrujillo13,Rivas11} 
and numerical simulations \cite{Sanders05,Denisov11,Rivas11,
Cattuto06,Shaebani12} confirm that the confinement of 
fluctuations in fluidized granular media produces spatial 
hydrodynamic inhomogeneities that are practically strong 
enough to induce substantial effective interactions between 
the intruders. Depending on the choice of parameter values, 
these forces can be either attractive or repulsive and may 
lead to a variety of collective behaviors 
\cite{Aumaitre01,Rivas11,GarciaTrujillo13}. Even a single 
asymmetric object may experience a self-force in a 
nonequilibrium state \cite{Buenzli09}. The interactions 
emerging in equilibrium state had also been studied by 
considering two bodies immersed in a solution of 
macromolecules \cite{Asakura54}. Note that we 
deal with the regime where the interparticle forces are 
mainly binary collisions. Investigation of the underlying 
mechanisms when durable frictional contacts dominate the 
behavior is beyond the scope of the present study.  

In this paper we first recall how the fluctuations of the 
hydrodynamic fields are spatially correlated in a randomly 
driven granular fluid. We derive the explicit form of the 
static structure factors, which reflect the spatial 
correlations of the medium. These quantities are then used 
to calculate the effective long-range interaction between 
two immobile big particles immersed in the medium. 
The method had been used in Ref.\cite{Shaebani12} to 
study the sign change and nonadditivity of the effective 
force, and to obtain the phase diagram of the transition 
between repulsive and attractive forces. Here, the main 
goal is to provide a detailed methodology for obtaining 
the structure factors and the fluctuation-induced force, 
as a recipe for further applications and extensions. We 
also investigate the effect of small changes in the 
boundary conditions on the effective force, by 
considering elongated intruders and changing their 
shape or orientation. The elongated particles are 
frequent and play important role in nature and industry. 
The interesting question arises whether fluctuation-induced 
forces may lead to orientational ordering and alignment 
of such particles. We investigate this question in a 
simple case of two elliptical intruders (with fixed 
centers) by varying their eccentricity and orientations 
to find the configuration where the intruders experience 
the maximum possible force. 

The paper is organized as follows: In Sec.~\ref{Theory-Section} 
we review the theory of randomly driven granular fluids and 
derive the structure factors, which are used in 
Sec.~\ref{EffectiveForce-Section} to calculate the effective 
force between the intruders with circular shapes. This part 
is the full exposition and expansion of the mode coupling 
calculations presented in \cite{Shaebani12}. The validity 
of the analytical method is established by means of discrete 
element method simulations in Sec.~\ref{NumericalResults-Section}. 
Section \ref{Shape-Section} is devoted to the investigation 
of the influence of intruder shape on the effective force, 
by considering elliptical particles. Finally, 
Sec.~\ref{Conclusion-Section} contains the discussion of the 
results and concluding remarks.

\section{Theory of a randomly driven granular system}
\label{Theory-Section}

In this section, we review the nonequilibrium steady state properties 
of a randomly driven granular system, and present the explicit form of 
the structure factors. A detailed description of the theory can be found, 
e.g., in Refs.~\cite{vanNoije99,vanNoije97,Gradenigo11}. Consider a system 
of $N$ hard spheres (or disks in two dimensions) of mass $m$ and radius 
$R$ with a Langevin-type equation of motion:
\begin{equation}
m \frac{d^2 {\bm r}_\text{i}(t)}{d t^2} = {\bm F}_\text{i}(t) + {\bm 
\xi}_\text{i}(t),
\label{Eq:Langevin}
\end{equation}
where ${\bm F}_\text{i}$ is the total force acting on particle $i$ 
due to binary inelastic collisions, and ${\bm \xi}_\text{i}$ is a 
random force arising from the coupling to an external heat bath. We 
assume that the random force is drawn from a distribution with zero 
mean, and with Cartesian components $\alpha$ and $\beta$ satisfying 
\begin{equation}
\langle \xi_{\text{i} \alpha}(t) \xi_{\text{j}\beta}(t') \rangle = 
2 \Gamma\delta_{\alpha \beta} \delta_{\text{ij}} \delta(t{-}t'). 
\label{Eq:NoiseCorr}
\end{equation}
Here, $\langle \cdot \cdot \cdot \rangle$ denotes averaging over the 
uncorrelated noise source, and $\Gamma$ reflects the noise strength. 
For the mean rate of energy gain of a single particle in $D$ dimensions 
one finds the following expression
\begin{equation}
\bigg\langle \frac{\partial \mathcal{E}}{\partial t} 
\bigg\rangle_{\!\!\text{gain}} = D \, \Gamma/m.
\label{Eq:EnergyGain}
\end{equation}
On the other hand, the mean-field rate of energy loss of a single 
particle by means of inelastic collisions is given by 
\cite{Enskog,vanNoije97}
\begin{equation}
\bigg\langle \frac{\partial \mathcal{E}}{\partial t} 
\bigg\rangle_{\!\!\text{loss}} = -\gamma \: (1{-}\alpha^2) \: \mathit{f}(n) 
\: T^{3/2},
\label{Eq:EnergyLoss}
\end{equation}
where $\gamma$ is a constant (which depends on $D$, $m$, and $R$), 
$\alpha$ is the normal restitution coefficient, and $\mathit{f}$ is 
a function of the number density $n$, given in two dimensions by 
\cite{Verlet82}
\begin{equation}
\mathit{f}(n){=} n \big(1{-}7\phi/16\big)\big/\big(1{-}\phi\big)^2,
\label{Eq:fn}
\end{equation}
with $\phi$ being the volume fraction. It can be seen from 
Eqs.~(\ref{Eq:EnergyGain}) and (\ref{Eq:EnergyLoss}) that the 
rate of temperature change is given by $\partial T / \partial t 
{=} D \Gamma/ m {-} \gamma (1{-}\alpha^2) \mathit{f}(n) T^{3/2}$, 
thus, the time dependence of temperature can be obtained by 
the following integration
\begin{equation}
\displaystyle \int_{T_{_0}}^T \displaystyle \frac{dT}{1 {-} 
\displaystyle \frac{m \gamma}{D \Gamma} \big(1{-}\alpha^2\big) 
\mathit{f}(n) T^{3/2}} = \displaystyle \int_{{_0}}^t 
\displaystyle \frac{D \Gamma}{m} dt.
\label{Eq:Tintegration}
\end{equation}
We introduce the scaled temperature $\widetilde T {\equiv} 
T/T_{_\text{MF}}$ (with $T_{_\text{MF}} {=} [D \Gamma/m\gamma 
(1{-}\alpha^2) \mathit{f}(n)]^{2/3}$ being the mean-field steady 
state temperature), and integrate Eq.~(\ref{Eq:Tintegration}) to 
arrive at the following implicit expression for the time evolution 
of the temperature
\begin{equation}
\ln \Big({\displaystyle\frac{1 {+} \displaystyle\sqrt{\widetilde T} {+} 
\widetilde T} {1 {-} 2 \displaystyle\sqrt{\widetilde T} 
{+} \widetilde T} }\Big)- 2{\sqrt 3} \arctan\Big({\displaystyle\frac{2 
\displaystyle\sqrt{\widetilde T} {+} 1}{\sqrt 3}}\Big) = \frac{3D 
\Gamma} {m T_{_\text{MF}}} \, t - A,
\label{Eq:Tevolution}
\end{equation}
where $A$ is a constant which depends on the initial temperature.
The system eventually reaches a nonequilibrium steady state, where 
the hydrodynamic fields $(T(\bm r,t),n(\bm r,t),\bm v(\bm r,t))$ 
fluctuate around their stationary values $(T_{\text{s}},n_{\text{s}}
({=}N/V),0)$, due to the stochastic nature of energy gain and loss 
($V$ is the volume of the $D${-}dimensional system). It has been 
proven \cite{vanNoije99} that the fluctuations in two dimensions 
are logarithmically divergent in the system size $L$, which results 
in a steady temperature higher than $T_{_\text{MF}}$ ($T_{\text{s}}
{-}T_{_\text{MF}} {\sim} \text{Ln}\:L$). The goal of the next 
subsection is to investigate the spatial fluctuations of the 
hydrodynamic fields, and their long-range correlations.

\subsection{Hydrodynamic fluctuations}
\label{HydroFluct-Subsection}

Assuming continuous (coarse-grained) hydrodynamic fields for a 
weakly inelastic fluid of hard particles, one can describe the 
system via standard hydrodynamic equations
\cite{vanNoije99,vanNoije00}
\begin{equation}
\begin{aligned}
\frac{\partial n(\bm r,t)}{\partial t} &= - {\bm \nabla} {\cdot} 
(n(t){\bm v}(t)),\\
\frac{\partial {\bm v}(\bm r,t)}{\partial t}&= -{\bm v} {\cdot} 
{\bm {\nabla v}} - \frac{1}{\rho} {{\bm \nabla} {\cdot} {\bm \Pi}},\\
\frac{\partial T(\bm r,t)}{\partial t} &= -{\bm v} {\cdot} {\bm 
\nabla} T - \frac{2}{D n}({\bm \nabla} {\cdot} {\bm J}+{\bm \Pi} {:} 
{\bm \nabla} {\bm v}) \\
&\hspace{17.5mm}{+} \bigg\langle \frac{\partial \mathcal{E}}
{\partial t} \bigg\rangle_{\!\!\text{gain}} {-} \bigg\langle 
\frac{\partial \mathcal{E}} {\partial t} \bigg\rangle_{\!\!\text{loss}},
\end{aligned}
\label{Eq:HydroEqs}
\end{equation}
where $\rho$, $\bm \Pi$, and $\bm J$ refer to the density, the 
pressure tensor, and the heat flux, respectively. The last equation 
is generalized by adding source and sink terms to account for the 
energy gain from the heat bath and the energy loss due to inelastic 
collisions. One can linearize Eqs.~(\ref{Eq:HydroEqs}) around the 
steady state homogeneous values $(T_{\text{s}},n_{\text{s}},0)$ 
to get the following equations \cite{vanNoije99}
\begin{equation}
\begin{aligned}
\frac{\partial}{\partial t} \delta n(\bm r,t) &{=} - n{\bm \nabla} 
{\cdot} {\bm v},\\ 
\frac{\partial}{\partial t} \delta \bm v(\bm r,t) &{=} 
-\frac{1}{\rho} \bm \nabla p {+} \eta \nabla^2 \bm v {+} 
(\eta{-}\eta_{_\ell}) {\bm {\nabla \nabla}} {\cdot} {\bm v} {+} 
{\bm \zeta}_1(\bm r,t),\\ 
\frac{\partial}{\partial t} \delta T(\bm r,t) &{=} \frac{2c}{Dn} 
\nabla^2 \delta T {-} \frac{2p}{Dn} {\bm \nabla} {\cdot} {\bm v} {-} 
\delta \bigg\langle\frac{\partial \mathcal{E}} {\partial t} 
\bigg\rangle_{\!\!\text{loss}} {+} {\zeta}_2(\bm r,t),
\end{aligned}
\label{Eq:LinearHydroEqs}
\end{equation}
with $p$, $c$, $\eta$, and $\eta_{_\ell}$ being the pressure, the 
heat conductivity, and the kinetic and longitudinal viscosities, 
respectively. Since the theory is valid for small inelasticities, 
$c$, $\eta$, and $\eta_{_\ell}$ can be approximated by the corresponding 
values in an elastic hard particle system, obtained from the Enskog 
theory \cite{Enskog}. The noise terms ${\bm \zeta}_1(\bm r,t)$ and 
${\bm \zeta}_2(\bm r,t)$ arise from (i) the external noise originating 
from the random force $\bm \xi(t)$ in Eq.~(\ref{Eq:Langevin}) 
\cite{vanNoije99}, and (ii) the internal fluctuations around thermal 
equilibrium, obtained from the fluctuation-dissipation theorem 
\cite{Landau59,vanNoije97}.

Next, using Fourier transforms, one replaces the hydrodynamic fluctuating 
fields $\delta F(\bm r,t)$ (including $n$, $T$, longitudinal velocity 
$v_{\!_\ell}$, and transverse velocity $v_t$) with $\delta F(\bm r,t) {=} 
\int \delta F(\bm k,t) e^{i \bm {k \cdot r}} d \bm k$ to get
\begin{equation}
\left(\begin{array}{c}
\!\!\frac{\partial}{\partial t} \delta n(\bm k,t) \\
\\
\!\!\frac{\partial}{\partial t} \delta T(\bm k,t) \\
\\
\!\!\frac{\partial}{\partial t} \delta v_{\!_\ell}(\bm k,t) \\
\\
\!\!\frac{\partial}{\partial t} \delta v_t(\bm k,t) 
\end{array}\!\!\right) {=}
-{\bm M(k)} \!\!
\left(\begin{array}{c}
\!\!\delta n(\bm k,t) \\
\!\!\delta T(\bm k,t) \\
\!\!\delta v_{\!_\ell}(\bm k,t) \\
\!\!\delta v_t(\bm k,t) 
\end{array}\!\!\right) {+}
\left(\begin{array}{c}
\!\!\zeta_{_n}(\bm k,t) \\
\!\!\zeta_{_T}(\bm k,t) \\
\!\!\zeta_v{_{\!_\ell}}(\bm k,t) \\
\!\!\zeta_v{_{\!_t}}(\bm k,t) 
\end{array}\!\!\right),
\label{Eq:FourierHydroEqs}
\end{equation}
with 
\begin{equation}
{\bm M(k)} {=}
\!\!\left(\begin{array}{cccc}
\!\!\!\!0 & 0 & \!ikn & \!\!\!0 \\
\\
\!\!\!\!\displaystyle\frac{(1{-}\alpha^2)\omega T}{D \; 
\mathit{f}(n)} \frac{d\mathit{f}(n)}{dn} 
& \displaystyle\frac{2ck^2}{D \, n}{+}\frac{3\omega(1{-} 
\alpha)^2}{2 \, D} & \!\displaystyle\frac{2ikp}{D 
\, n} & \!\!\!0 \\
\\
\!\!\!\!\displaystyle\frac{ik}{n}\left(\displaystyle\frac{\partial 
p}{\partial \rho}\right)_{\!T} & \displaystyle\frac{ikp}{\rho T} 
& \!\eta_{_\ell} k^2 & \!\!\!0 \\
\\
\!\!\!\!0 & 0 & \!0 & \!\!\!\eta k^2 
\end{array}\!\!\!\!\right)\!\!.
\label{Eq:Mmatrix}
\end{equation}
Here, $\omega {\sim} \mathit{f}(n) \sqrt{T}$ is the 
collision frequency obtained from the Enskog theory. The 
component $\zeta_{_n}(\bm k,t)$ of the white noise vector 
$\bm \zeta(\bm k,t)$ in Eq.~(\ref{Eq:FourierHydroEqs}) 
is indeed zero, and the rest of the components (i.e.\ 
$\zeta_{_T}(\bm k,t)$, $\zeta_v{_{\!_\ell}}(\bm k,t)$, and 
$\zeta_v{_{\!_t}}(\bm k,t)$) are Gaussian with correlations
\begin{equation}
\langle \zeta_\text{i}(\bm k,t) \zeta_\text{j}(-\bm k, t') 
\rangle = V C_{_\text{ij}}(k) \delta(t{-}t'). 
\label{Eq:zetaNoise}
\end{equation}
By taking Fourier transforms of ${\bm \zeta}_1(\bm r,t)$ and 
${\bm \zeta}_2(\bm r,t)$, the nonzero elements of the matrix 
${\bm C}(k)$ are determined as follows \cite{vanNoije99}
\begin{eqnarray}
\begin{aligned}
C_{_\text{22}}(k) &{=} \displaystyle\frac{8T\Gamma}{D \rho}{+}
\displaystyle\frac{8cT^2}{D^2n^2}k^2,\\
\\
C_{_\text{33}}(k) &{=} \displaystyle\frac{2n\Gamma}{\rho^2}
{+}\displaystyle\frac{2\eta_{_\ell} T}{\rho}k^2,\\
\\
C_{_\text{44}}(k) &{=} \displaystyle\frac{2n\Gamma}{\rho^2}
{+}\displaystyle\frac{2\eta T}{\rho}k^2.
\end{aligned}
\label{Eq:C22C33C44}
\end{eqnarray}

Once the hydrodynamic fluctuations are determined, one can 
calculate the structure factors $S_{_{ab}}(k)$, which 
are indeed the Fourier transforms of the spatial correlation 
functions. The structure factors in the nonequilibrium steady 
state are given by
\begin{equation}
S_{_{ab}}(k){=}\lim_{t\to\infty}\frac{1}{V}\langle \delta 
a(\bm k,t) \, \delta b(-\bm k,t) \rangle.
\label{Eq:StructDef}
\end{equation}
Let us denote $\bm S(k)$ as a $4 \times 4$ matrix with elements 
labeled by $n$, $T$, $v_{\!_\ell}$, and $v_t$: 
\begin{equation}
\bm S(k) {=}
\!\left(\begin{array}{cccc}
\!\!S_{nn}(k) & S_{nT}(k) & S_{nv_{\!_\ell}}(k) & S_{nv_{\!_t}}(k) \\
\\
\!\!S_{T\!n}(k) & S_{TT}(k) & S_{Tv_{\!_\ell}}(k) & S_{T\!v_{\!_t}}(k) \\
\\
\!\!S_{v_{\!_\ell}\!n}(k) & S_{v_{\!_\ell}\!T}(k) & S_{v_{\!_\ell\!}
v_{\!_\ell}}(k) & S_{v_{\!_\ell}\!v_{\!_t}}(k) \\
\\
\!\!S_{v_{\!_t}\!n}(k) & S_{v_{\!_t}\!T}(k) & S_{v_{\!_t}
\!v_{\!_\ell}}(k) & S_{v_{\!_t}\!v_{\!_t}}(k)
\end{array}\!\!\right)\!.
\label{Eq:Smatrix}
\end{equation}
\vspace{5mm}

By integrating Eqs.~(\ref{Eq:FourierHydroEqs}) and using 
Eq.~(\ref{Eq:zetaNoise}) one arrives at the following equation 
for the time evolution of $\bm S(k)$
\begin{equation}
\frac{\partial}{\partial t}{\bm S(k)} = -{\bm M(k)} {\bm S(k)} 
- {\bm S(k)}{\bm M^T(-k)} + \bm C(k),
\label{Eq:Sevolution}
\end{equation}
where $\bm M^T$ is the transpose of $\bm M$. Setting the left-hand 
side to zero, the steady state values of the structure factors can 
be calculated. Among the elements of $\bm S(k)$, we are interested 
in $S_{nn}(k)$ and $S_{nT}(k)$ obtained as
\begin{widetext}
\begin{equation}
S_{nn}(k) = \frac{M_{_{13}} \big(M_{_{22}} M_{_{32}}^2 C_{_{22}} + 
M_{_{32}}^2 M_{_{33}} C_{_{22}} - M_{_{22}}^3 C_{_{33}} + M_{_{13}} 
M_{_{22}} M_{_{31}} C_{_{33}} - M_{_{13}} M_{_{21}} M_{_{32}} C_{_{33}} 
- M_{_{22}}^2 M_{_{33}} C_{_{33}} \big)}{2 \big(M_{_{22}} M_{_{31}} - 
M_{_{21}} M_{_{32}} \big) \bigg( \big(M_{_{22}} + M_{_{33}} \big) 
\big(M_{_{23}} M_{_{32}} - M_{_{22}} M_{_{33}} \big) + M_{_{13}} 
\big(M_{_{21}} M_{_{32}} + M_{_{31}} M_{_{33}} \big)\bigg)},
\label{Eq:SnnFull}
\end{equation}
and
\begin{equation}
S_{nT}(k) = \frac{M_{_{13}} \bigg(M_{_{31}} \big(-M_{_{32}} M_{_{33}} 
C_{_{22}} + M_{_{22}} M_{_{23}} C_{_{33}} \big) + M_{_{21}} 
\big(-M_{_{32}}^2 C_{_{22}} - M_{_{23}} M_{_{32}} C_{_{33}} + M_{_{22}} 
\big(M_{_{22}} + M_{_{33}} \big) C_{_{33}} \big) \bigg)}{2\big(M_{_{22}} 
M_{_{31}} - M_{_{21}} M_{_{32}} \big) \bigg(\big(M_{_{22}} + M_{_{33}} 
\big) \big(M_{_{23}} M_{_{32}} - M_{_{22}} M_{_{33}} \big) + M_{_{13}} 
\big(M_{_{21}} M_{_{32}} + M_{_{31}} M_{_{33}} \big)\bigg)}.
\label{Eq:SnTFull}
\end{equation}
Because of the dominant contribution of small wave numbers $k$, one can 
approximate $S_{nn}(k)$ and $S_{nT}(k)$ by their leading terms 
($\mathcal{O}(1/k^2)$) which leads to
\begin{equation}
S_{nn}(k) \simeq \frac{-27 \omega^2 T^2 n^3 \rho (1{-}\alpha^2)^2}{4D 
\bigg[\bigg(\displaystyle\frac{2np}{\mathit{f}(n)} \displaystyle\frac{d
\mathit{f}(n)}{dn} - 3 \rho (\displaystyle\frac{\partial p}{\partial 
\rho})_{\!_T} \bigg) \bigg(6 p^2 + \displaystyle\frac{2Dn^2pT}{\mathit{f}(n)}
\displaystyle\frac{d\mathit{f}(n)}{dn} +\displaystyle\frac{9 
(1{-}\alpha^2) n \eta_{_\ell} \omega \rho T}{2} \bigg)\bigg]}
\displaystyle\frac{1}{k^2},
\label{Eq:SnnApp}
\end{equation}
and
\begin{equation}
S_{nT}(k) \simeq \frac{9 \omega^2 T^3 n^3 \rho (1{-}\alpha^2)^2 
\displaystyle\frac{d\mathit{f}(n)}{dn}}{2 D \mathit{f}(n) 
\bigg[\bigg(\displaystyle\frac{2np}{\mathit{f}(n)} \displaystyle\frac{d
\mathit{f}(n)}{dn} - 3 \rho (\displaystyle\frac{\partial p}{\partial 
\rho})_{\!_T} \bigg) \bigg(6 p^2 + \displaystyle\frac{2Dn^2pT}{\mathit{f}(n)}
\displaystyle\frac{d\mathit{f}(n)}{dn} +\displaystyle\frac{9 
(1{-}\alpha^2) n \eta_{_\ell} \omega \rho T}{2} \bigg)\bigg]}
\displaystyle\frac{1}{k^2}.
\label{Eq:SnTApp}
\end{equation}
\end{widetext}
Note that the validity of the mode coupling theory presented in 
this section is restricted to the long wavelength range, where 
$k_\text{max} {<} \text{min}[ 2\pi/l^* , \pi/R]$ (with $l^*$ 
being the mean free path and $R$ the particle radius). The wave 
numbers are limited also by the system size, so that $2\pi/L 
{\leq} k_\text{min}$. It is also notable that the approximate 
value of $S_{nn}(k)$ [Eq.~(\ref{Eq:SnnApp})] is indeed 
independent of the noise strength, while $S_{nT}(k)$ 
[Eq.~(\ref{Eq:SnTApp})] is proportional to $\Gamma^{2/3}$. 

\section{Long range interactions between intruder particles}
\label{EffectiveForce-Section}

The fluctuating hydrodynamic fields in the nonequilibrium 
steady state of a uniformly driven medium are spatially 
homogeneous, as discussed in Sec.~\ref{Theory-Section}. 
However, in the presence of large immobile particles 
immersed in the granular fluid, the fluctuation spectrum 
would be modified according to the new boundary conditions, 
resulting in inhomogeneous fields. In the following, we 
analyse the pressure fluctuations $\delta p(\bm r,t)$ (in 
two dimensions for simplicity) and show how geometric 
constraints lead to average pressure difference around 
the intruder particles. Let us start with the Verlet-Levesque 
equation of state for a hard disk system \cite{Verlet82}
\begin{equation}
p(n,T) = g(n) T,
\label{Eq:Verlet-Levesque}
\end{equation}
with $g(n){=}n(1+\phi^2/8)/(1-\phi)^2$, where $\phi{=}\pi 
R^2 n$ is the volume fraction. We expand $p(n,T)$ up to 
second order around the steady state values $(n_{\text{s}},
T_{\text{s}})$ 
\begin{equation}
\begin{aligned}
\delta &p(\bm r,t) {=} p(\bm r,t) {-} p(n_{\text{s}},T_{\text{s}}) 
\simeq \frac{1}{2} T_{\text{s}} \frac{d^2g}{dn^2}\big|_{_{n{=} 
n_{\text{s}}}} \hspace{-2mm}\big( \delta n(\bm r,t) \big)^2+\\
&g(n_{\text{s}})\delta T(\bm r,t) {+} T_{\text{s}} 
\frac{dg}{dn}\big|_{_{n{=}n_{\text{s}}}} \hspace{-3mm}\delta n(\bm r,t) 
{+} \frac{dg}{dn}\big|_{_{n{=}n_{\text{s}}}} 
\hspace{-3mm}\delta n(\bm r,t) \delta T(\bm r,t).
\end{aligned}
\label{Eq:PressureExp1}
\end{equation}
The statistical average of $\delta p(\bm r,t)$ over the random 
noise source is then given by 
\begin{equation}
\begin{aligned}
&\langle \delta p(\bm r,t) \rangle {=}\\
&\frac{dg}{dn}\big|_{_{n{=}n_{\text{s}}}} 
\hspace{-1mm} \langle \delta n(\bm r,t) \delta T(\bm r,t) \rangle {+} 
\frac{1}{2} T_{\text{s}} \frac{d^2g}{dn^2}\big|_{_{n{=}n_{\text{s}}}} 
\hspace{-1mm}\langle  \big( \delta n(\bm r,t) \big)^2 \rangle,
\end{aligned}
\label{Eq:PressureExp2}
\end{equation}
and finally, by employing the Fourier transforms of the fluctuations, 
one obtains the local pressure fluctuations as a function of the static 
structure factors 
\begin{equation}
\delta p(k) {\sim} \int \bigg( \frac{dg}{dn}\big|_{_{n{=}n_{\text{s}}}} 
\hspace{-1mm} S_{nT}(k) {+} \frac{1}{2} T_{\text{s}} 
\frac{d^2g}{dn^2}\big|_{_{n{=}n_{\text{s}}}} \hspace{-1mm} S_{nn}(k)\bigg) dk.
\label{Eq:PressureExp3}
\end{equation}
The upper bound of integration is $k_\text{max} {=} 2\pi / \text{max}[l^* , 
2R]$ (as discussed in the previous section) to ensure that the hydrodynamic 
description is valid. The lower bound of the allowed wave numbers is 
restricted by the geometric constraints. Hence, the presence of intruder 
particles would change the local pressure fluctuations, leading to a net 
pressure difference around each intruder. As a result, the immersed 
particles receive forces which can be interpreted as effective 
long-range interactions between them. 

\begin{figure}[b]
\centering
\includegraphics[scale=0.3,angle=0]{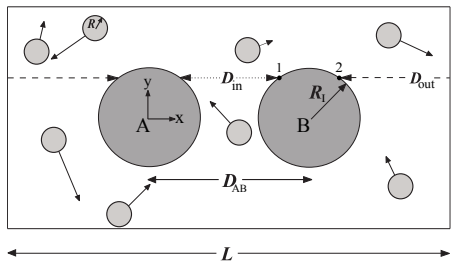}
\caption{Sketch of the two dimensional system.}
\label{Fig1}
\end{figure}

As an illustrating example, we consider a system with two fixed intruders 
A and B with the same radius $R_{_I}$ (see Fig.~\ref{Fig1}) and compare 
the pressure fluctuations at two points 1 and 2 which have the same $y$ 
coordinates and located on opposite sides of particle B. While the 
spectrum along $y$ direction at both points is restricted to the same 
limits, the allowed wave vectors in $x$ direction are confined to the 
horizontal distances $D_{_\text{in}}$ and $D_{_\text{out}}$ given by
\begin{eqnarray}
\begin{aligned}
D\!_{_\text{in}}\!(y) &{=}D\!_{_\text{AB}}\!\!-\!
2\displaystyle\sqrt{R_{_I}^2\!\!-\!\!y^2},\\
D\!_{_\text{out}}\!(y) &{=}L\!-\!D\!_{_\text{AB}}
\!\!\!-\!\!2\displaystyle\sqrt{R_{_I}^2\!\!-\!\!y^2}.
\label{Eq:Din-Dout}
\end{aligned}
\end{eqnarray}
Therefore, the local fluctuation of pressure is different at these 
two points, which gives rise to an average pressure difference
\begin{equation}
\Delta p (y) {=}\bigg( \int_{2\pi\!/\!D\!_{_\text{in}}\!\!(y)}^
{k_{_\text{max}}}\!\!\!\!\!\!\!\!\! dk_{_x} \!\!-\!\!
\int_{2\pi\!/\!D\!_{_\text{out}}\!(y)}^{k_{_\text{max}}}
\!\!\!\!\!\!\!\!\! dk_{_x} \bigg) \!\!\int_{k_{_\text{min}}\!\!(y)}^
{k_{_\text{max}}}\!\!\!\!\!\!\! dk_{_y} \:g(k_{_x},k_{_y}),
\label{Eq:PressureDiff}
\end{equation}
where $g(k_{_x},k_{_y})$ is the integrand in Eq.~(\ref{Eq:PressureExp3}). 
The total effective force $F$ acting on particle B can be obtained 
by integrating the pressure difference over the surface of the particle
\begin{equation}
F = \int_{-R_{_I}}^{R_{_I}} \Delta p (y) dy.
\label{Eq:EffectiveForce}
\end{equation}
The magnitude and sign of the resulting force depends on $\Delta 
p (y)$. In general, one should solve $\Delta p {=} 0$ to get the 
crossover point between attractive and repulsive forces. However, 
using the approximate values of the structure factors 
$S_\text{nn} (k)$ and $S_\text{nT} (k)$ [Eqs.~(\ref{Eq:SnnApp}) 
and (\ref{Eq:SnTApp})], i.e.\ only the leading terms 
($\mathcal{O}(1/k^2)$), the integrand in Eq.~(\ref{Eq:PressureDiff}) 
can be written as $g(k){\sim}\frac{f(n,\alpha)}{k^2}$. Therefore, the 
function $f(n,\alpha)$ can be taken out from the integral, and the 
transition point is approximately obtained by setting $f(n,\alpha)$ 
to zero, which leads e.g.\ to $\phi_c{\sim}0.57$ for $\alpha{=}0.8$. 
The transition point in general is determined by the restitution 
coefficient, the density, and the distance between the intruders 
\cite{Shaebani12}, thus, one should use the full expressions 
of the structure factors [Eqs.~(\ref{Eq:SnnFull}) and 
(\ref{Eq:SnTFull})]. In the next section, we compare the mode 
coupling predictions with the results obtained from the simulations. 

\section{Simulation details}
\label{NumericalResults-Section}

\subsection{The model}
\label{Model-Subsection}

We consider a two-dimensional granular gas by means of discrete 
element method simulations. The system consists of $N{=}3000$ 
identical rigid disks of radius $R$ and mass $m$ interacting via
inelastic collisions with the normal coefficient of restitution 
$\alpha{=}0.8$ for all collisions. In order to provide a spatially 
homogeneous state and exclude the undesired effects of side walls, 
periodic boundary conditions are applied in both directions of the 
square-shaped system of length $L {=} 200 R$. Two immobile rigid 
intruder particles of radius $R_{_I}{=}10 R$ and infinite mass and 
moment of inertia are immersed in the granular gas bed, separated 
by a distance $D_{\!_\text{AB}}{=}30R$ (see Fig.~\ref{Fig1}). The 
interaction with the heat bath is modeled in a similar way as in 
Refs.~\cite{Peng98,vanNoije99,Cattuto06,Shaebani12}, where the 
momentum of each particle is perturbed instantaneously at each 
time step $\Delta t$ by a random amount taken from a Gaussian 
distribution with zero mean and a given variance $\sigma$. We 
note that the assumption of the Gaussian white noise is satisfied 
in the limit $\sigma {\to} 0$, and providing that the mean free 
time $\tau^*$ of the granular gas remains much larger than $\Delta t$.
Another point is that the total linear momentum of the system is 
not necessarily conserved in our simulations, since we do not set 
the sum of noise vectors to zero at each time step. However, the 
deviation decreases as $1/N$ and vanishes in the limit of large $N$.

\begin{figure}[t]
\centering
\includegraphics[scale=0.40,angle=0]{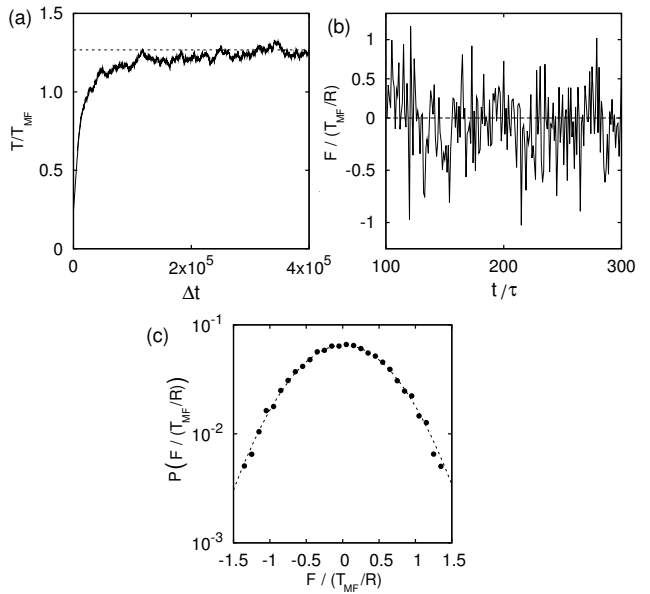}
\caption{(a) Time evolution of temperature $T$ (scaled by 
$T_{_\text{MF}}$) starting from a random initial state. 
The dashed line indicates the final stationary level. (b) 
A typical sequence of the fluctuation-induced forces $F$ 
scaled by $T_{_\text{MF}}/R$. (c) The probability distribution 
of $F$ collected over $10^8$ time steps. The dashed line shows 
the best fit with a Gaussian distribution centered on $0.015 
T_{_\text{MF}}/R$.}
\label{Fig2}
\end{figure}

\subsection{Results}
\label{Results-Subsection}

\begin{figure}[t]
\centering
\includegraphics[scale=0.40,angle=0]{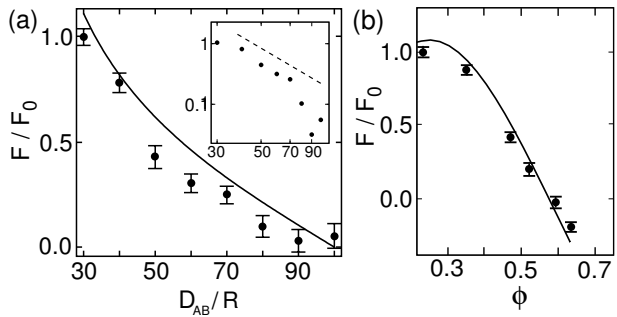}
\caption{The effective force $F$ scales by $F_{_0}$ 
in terms of (a) the distance $D\!_{_\text{AB}}$ between the intruders, 
and (b) the volume fraction $\phi$. The solid lines correspond to 
the mode coupling results discussed in Sec.~\ref{EffectiveForce-Section}. 
Inset: Same plot as in (a) but on a log-log scale. The dashed line 
${\sim} D_{\!_\text{AB}}^{-2}$ is a guide to the eye.}
\label{Fig3}
\end{figure}

Starting from a random initial state, the driven gas finally converges to 
a nonequilibrium steady state, as shown in Fig.~\ref{Fig2}(a). Then 
after the system relaxes, we investigate the effective interaction 
between the two fixed intruders. The total force exerted by the 
small beads on each intruder during the time interval $\tau$ 
($\tau \!\!\sim\!\! 250$ collisions per particle) is measured. 
Figure \ref{Fig2}(b) displays the fluctuating nature of the resulting 
consecutive forces, therefore, this quantity is averaged over more than 
$10^4$ time intervals $\tau$ to suppress the observed large fluctuations. 
The probability distribution of the data measured along the $x$ 
axis, shown in Fig.~\ref{Fig2}(c), can be well fitted by a Gaussian 
\cite{Bartolo02} with the average value $0.015 T\!_{_\text{MF}}/R 
\!\equiv\!F_{_0}$ and the standard deviation $\sigma {=} 0.612 
T\!_{_\text{MF}}/R$. The fluctuations are about two orders of 
magnitude larger than the average value, indicating that very 
long simulations are required to measure the forces with relatively 
small numerical errors. Repeating a similar procedure along the 
$y$ axis leads to almost zero force within the accuracy of our 
measurements. $F_{_0}$ can be considered as the magnitude of the 
effective long-range force acting between intruders A and B, which 
is repulsive in this case. 

Next we compare the simulation results with the mode coupling 
predictions obtained from 
Eqs.~(\ref{Eq:PressureExp3})-(\ref{Eq:EffectiveForce}) using the 
full expressions of the structure factors [Eqs.~(\ref{Eq:SnnFull}) 
and (\ref{Eq:SnTFull})]. In Fig.~\ref{Fig3}(a), the force is 
shown as a function of the distance between the intruders. The 
interaction is weakened with increasing the distance, and vanishes 
at $D\!_{_\text{AB}}{=}L/2$ as expected from the symmetry. By 
varying the density of the small grains, it is shown in 
Fig.~\ref{Fig3}(b) that the sign of the force changes as the 
volume fraction exceeds a threshold value $\phi_{_c} \sim 0.57$. 
The analytical results are also shown in Fig.~\ref{Fig3} (solid 
lines). The agreement is satisfactory, however, the mode coupling 
method slightly overestimates the force. The deviation can be 
attributed to the fact that the fluctuations on the opposite 
sides of the intruder are indeed correlated, which is not taken 
into account in the mode coupling calculations 
\cite{Cattuto06,Shaebani12}. We also point out that using the 
approximate values of the structure factors with leading terms 
of order $1/k^2$ [Eqs.~(\ref{Eq:SnnApp}) and (\ref{Eq:SnTApp})] 
would lead to qualitatively similar results but with more than 
$10\%$ errors compared with the full expressions 
[Eqs.~(\ref{Eq:SnnFull}) and (\ref{Eq:SnTFull})]. A more detailed 
study of the transition behavior, and further comparison between 
theoretical and simulation results can be found in \cite{Shaebani12}.

\section{Influence of shape}
\label{Shape-Section}

\begin{figure}[b]
\centering
\includegraphics[scale=0.65,angle=0]{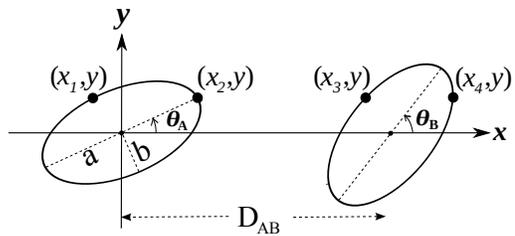}
\caption{Schematic picture depicting the two elliptical 
intruders and their orientations.}
\label{Fig4}
\end{figure}

As demonstrated in the previous sections, the fluctuation-induced 
force originates from the confinement of the fluctuation spectrum 
due to geometric constraints imposed by intruders. Two questions 
arise: How sensitive is the force to the tiny changes in the boundary 
conditions induced by changing the shape of the intruders? and is 
the fluctuation-induced interaction able to align elongated particles 
or create patterns of orientational order? In this section we modify 
the intruders shape and use the mode coupling method to analytically 
calculate the corresponding effective forces between two elongated 
intruders. Numerical simulations to study this effect would be extremely 
time consuming and are beyond the scope of the present investigation. 
Instead of circular intruders, here we choose elliptical particles 
with radii $a$ and $b$ ($b{<}a$), as shown in Fig.~\ref{Fig4}. By defining 
\begin{eqnarray}
\begin{aligned}
c_{_1} &{=} ( b^2 \text{cos}^2\theta_{\!_A} {+} a^2 \text{sin}^2\theta_{\!_A}) 
/ a^2 b^2,\\ 
c_{_2} &{=} ( b^2 \text{sin}^2\theta_{\!_A} {+} a^2 \text{cos}^2\theta_{\!_A}) 
/ a^2 b^2,\\ 
c_{_3} &{=} (b^2{-}a^2) \text{sin}\theta_{\!_A} \text{cos}\theta_{\!_A} 
/ a^2 b^2,\\ 
c_{_4} &{=} ( b^2 \text{cos}^2\theta_{\!_B} {+} a^2 \text{sin}^2\theta_{\!_B}) 
/ a^2 b^2,\\ 
c_{_5} &{=} ( b^2 \text{sin}^2\theta_{\!_B} {+} a^2 \text{cos}^2\theta_{\!_B}) 
/ a^2 b^2,\\ 
c_{_6} &{=} (b^2{-}a^2) \text{sin}\theta_{\!_B} \text{cos}\theta_{\!_B} 
/ a^2 b^2,
\end{aligned}
\end{eqnarray}
one finds the distances $D\!_{_\text{in}}$ and $D\!_{_\text{out}}$ 
at a given $y$ coordinate as
\begin{eqnarray}
\begin{aligned}
&D\!_{_\text{in}}\!(y) {=} x_{_3} {-} x_{_2},\\
&D\!_{_\text{out}}\!(y) {=} L {-} x_{_4} {+} x_{_1},
\end{aligned}
\end{eqnarray}
with
\begin{eqnarray}
\begin{aligned}
x_{_1} &{=}-\frac{c_{_3}}{c_{_1}}y-\displaystyle\sqrt{\frac{c_{_1}+
(c_{_3}^2-c_{_1}c_{_2})y^2}{c_{_1}^2}},\\
x_{_2} &{=}-\frac{c_{_3}}{c_{_1}}y+\displaystyle\sqrt{\frac{c_{_1}+
(c_{_3}^2-c_{_1}c_{_2})y^2}{c_{_1}^2}},\\
x_{_3} &{=} D\!_{_\text{AB}} -\frac{c_{_6}}{c_{_4}}y-
\displaystyle\sqrt{\frac{c_{_4}+
(c_{_6}^2-c_{_4}c_{_5})y^2}{c_{_4}^2}},\\
x_{_4} &{=} D\!_{_\text{AB}} -\frac{c_{_6}}{c_{_4}}y+
\displaystyle\sqrt{\frac{c_{_4}+
(c_{_6}^2-c_{_4}c_{_5})y^2}{c_{_4}^2}}.
\end{aligned}
\end{eqnarray}
Thus, similar to Eq.~(\ref{Eq:PressureDiff}), the 
average pressure difference reads
\begin{equation}
\Delta p (y) {=} \int_{2\pi\!\big/\!\big(x_{_3}(y){-}x_{_2}(y)\big)}^
{2\pi\!\big/\!\big(L{-}x_{_4}(y){+}x_{_1}(y)\big)}\!\!\! dk_{_x} \!\!\int_{k_{_\text{min}}\!\!(y)}^
{k_{_\text{max}}}\!\!\! dk_{_y} \:g(k_{_x},k_{_y}).
\label{Eq:PressureDiffEllipses}
\end{equation}

\begin{figure}[t]
\centering
\includegraphics[scale=0.8,angle=0]{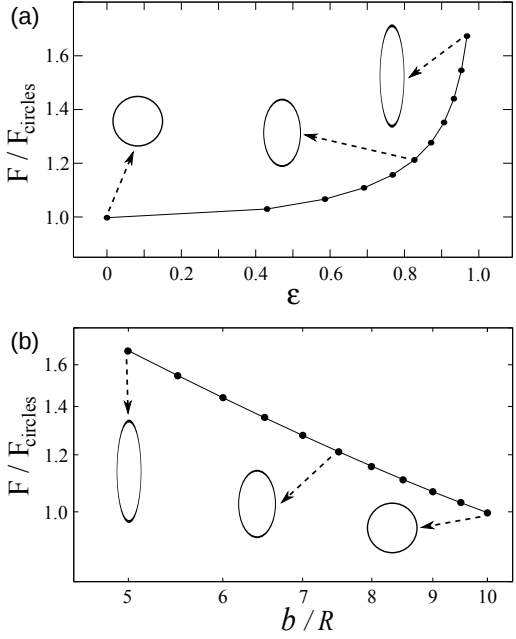}
\caption{The effective force between two ellipses in terms of (a) the 
eccentricity $\epsilon$, and (b) the small radius of the ellipses $b$. 
The distance between the centers of the intruders is fixed at 
$D\!_{_\text{AB}} {=} 30R$, and the forces are scaled by the force 
between two circles.}
\label{Fig5}
\end{figure}

\begin{figure}[b]
\centering
\includegraphics[scale=0.3,angle=0]{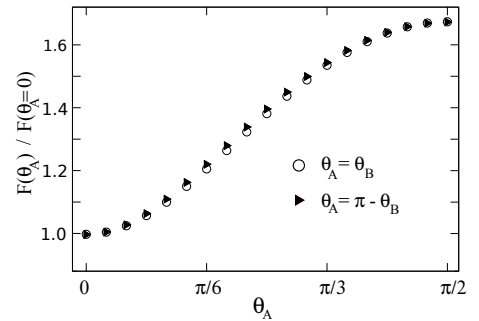}
\caption{The effective force between the ellipses vs.\ the angle 
$\theta_{\!_A}$. The forces are scaled by the force at 
$\theta_{\!_A}{=}0$.}
\label{Fig6}
\end{figure}

\begin{figure}[t]
\centering
\includegraphics[scale=0.7,angle=0]{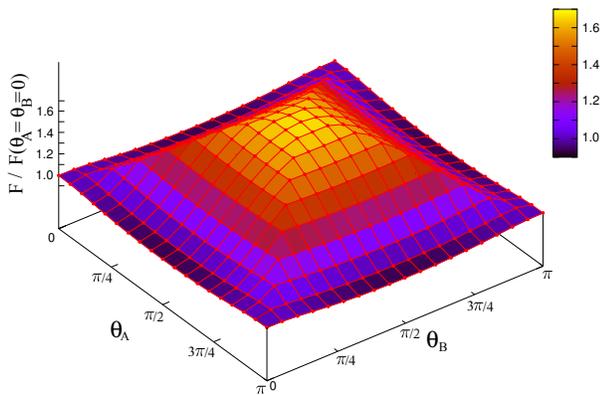}
\caption{(color online) The force between two ellipses in terms of 
the angles $\theta_{\!_A}$ and $\theta_{\!_B}$. The color intensity 
reflects the magnitude of $F$.}
\label{Fig7}
\end{figure}

Let us first fix the orientation of the ellipses at $\theta_{\!_A} {=} 
\theta_{\!_B} {=} \pi {/} 2$ and the distance between their centers 
at $D\!_{_\text{AB}} {=} 30R$. We modify the shape of intruders by 
changing the ratio between the ellipse radii, $a{/}b$, while its area 
is constant, i.e.\ $\pi a b {=} 100 \pi R^2 ({=} \pi R_{\!_I}^2)$. 
The deviation of the shape from being circular can be characterized 
with the eccentricity parameter defined as
\begin{equation}
\epsilon = \sqrt{1-\displaystyle\frac{b^2}{a^2}}.
\label{Eq:eccentricity}
\end{equation}
Using the same procedure as in Sec.~\ref{EffectiveForce-Section} we 
calculate the effective force between two ellipses as $F{=}\int_{-a}^a 
\Delta p (y) dy$. Figure \ref{Fig5} shows that the force considerably 
increases as the shape further deviates from being circular. The 
behavior of $F$ results from two competing effects: first, 
increasing $\epsilon$ causes lower pressure difference because 
$D\!_{_\text{out}}{/}D\!_{_\text{in}}$ decreases on average; second, 
it also increases the integration interval $[-a,a]$, which 
increases the force. The latter effect seems to be more important 
so that the force monotonically increases with $\epsilon$.

Next we investigate the influence of the orientations of the ellipses 
on the fluctuation-induced force. One expects that the eccentricity 
of the elliptical intruders also influences the force behavior when 
the orientation changes. Only in the limit of circular shapes the 
orientation plays no role anymore. Therefore, we fix the shape by 
choosing $a {=} 2b {=} 10 R$ and vary the angles $\theta_{\!_A}$ 
and $\theta_{\!_B}$. Let us first consider two cases 
$\theta_{\!_A}{=}\theta_{\!_B}$ and $\theta_{\!_A}{=} 
\pi{-}\theta_{\!_B}$ for simplicity, thus, there remains only 
one rotational degree of freedom. The force is given by 
\begin{equation}
F{=}\int_{-y_{_\text{max}}}^{y_{_\text{max}}} \Delta p (y) dy, 
\label{Eq:ForceEllipse}
\end{equation}
where the interval [$-y_{_\text{max}},y_{_\text{max}}$] is the range 
of $y$ coordinate covered by both intruders. In the case of 
$\theta_A {\neq} \theta_B$, each intruder covers a different range 
of $y$ coordinate, leading to a different value of $y_{_\text{max}}$. 
Thus, in general, we choose the smaller value of $y_{_\text{max}}$ 
to calculate the integral, i.e.\ only the area where two elliptical 
intruders face each other is taken into account
\begin{equation}
y_{_\text{max}} = \text{min} \Big( 
\displaystyle\frac{c_{_1}}{\displaystyle\sqrt{c_{_1}^2 c_{_2} - c_{_3}^2c_{_1}}},
\displaystyle\frac{c_{_4}}{\displaystyle\sqrt{c_{_4}^2 c_{_5} - c_{_6}^2c_{_4}}} 
\Big).
\end{equation}
With increasing the angle $\theta_{\!_A}$, on the one hand the average 
ratio $D\!_{_\text{out}}{/}D\!_{_\text{in}}$ decreases which weakens 
the pressure difference between the gap and outside regions. On the 
other hand, the integration interval $[-y_{_\text{max}},y_{_\text{max}}]$ 
is extended which increases the force. Figure \ref{Fig6} shows that 
the intruders exert stronger forces on each other in the case 
$\theta_{\!_A}{=}\theta_{\!_B}{=}\pi/2$ (maximum area, minimum 
$\Delta p$) compared with $\theta_{\!_A}{=} \theta_{\!_B}{=}0$ 
(minimum area, maximum $\Delta p$). One can also see that the 
interaction in the case of $\theta_{\!_A} {=} \pi {-} \theta_{\!_B}$ 
is slightly stronger than the case $\theta_{\!_A} {=} \theta_{\!_B}$ 
(except at $\theta_{\!_A} {=} 0, \pi/2$). The results of the general 
case of $\theta_{\!_A}{\neq}\theta_{\!_B}$ are shown in 
Fig.~\ref{Fig7}. The configuration with $\theta_{\!_A}{=} 
\theta_{\!_B}{=}\pi/2$ produces the strongest interaction, while 
the lowest force is obtained if $|\theta_{\!_A}{-}\theta_{\!_B}|
{=}\pi/2$ and $\theta_{\!_A}{=}\pi/2$ (or $\theta_{\!_B}{=}\pi/2)$. 
Therefore, the maximum fluctuation-induced force is obtained when 
the two ellipses are aligned along the $y$ axis. This result, 
however, can not be trivially extended to more complicated 
situations, and a comprehensive study is required to elucidate 
the role of key parameters such as gas density and restitution 
coefficient, as well as the number, position, and orientation 
of the intruders in a multi-body configuration.

\section{Discussion and Conclusion}
\label{Conclusion-Section}
We reviewed the detailed calculations of the mode coupling method to 
describe the nonequilibrium steady state of a randomly driven granular 
system. When the existence of arising long-range hydrodynamic 
correlations (reflected in the structure factors) is accompanied by 
geometric constraints, effective long-range forces appear. Using 
the pair correlation functions in the mode coupling calculations 
leads to a reasonable estimation of the force between two intruders. 
It is notable that the hydrodynamic correlations become stronger in 
multi-intruder configurations or binary mixtures, which necessitates 
the usage of triplet or higher order structure factors \cite{Shaebani12,
Dijkstra06}. The validity of the hydrodynamic description is also 
limited to large restitution coefficients. The steady state of an 
inelastic granular fluid is inhomogeneous, resulting in a length 
scale $l_i$ on which the macroscopic hydrodynamic fields vary. 
Decreasing the restitution coefficient would increase the gradient 
of the inhomogeneities, i.e.\ decreases the length scale $l_i$. The 
hydrodynamic description of the system holds only when $l_i$ is 
well separated from the mean-free path $l^*$ of the granular gas, 
and this happens only for large restitution coefficients.  

Another point is that the analytical results presented 
in this work are restricted to two dimensions but the procedure can 
be extended to three dimensions e.g.\ by considering a suitable 
equation of state for hard sphere systems, and calculating the force 
exerted on the surface of three-dimensional obstacles. Nevertheless, 
the behavior of pressure fluctuations is different in two and three 
dimensions (it behaves as $1/r$ in 3D while diverges logarithmically 
as $\text{ln}(L/r)$ in 2D). Thus, the force is independent of 
(logarithmically dependent on) the system size in three (two) 
dimensions \cite{Shaebani12}. Indeed, the long-range behavior of the 
correlations and the dimensional dependence of the effective force 
originate from the behavior of the inverse of the Laplacian in 
the hydrodynamic equations \cite{vanNoije99}. 

Finally, we have shown that changing the shape or orientation 
of the intruders influences the effective force in two different 
ways: (i) by modifying the gap length (along the $x$-direction), 
which affects the pressure difference around the intruders, (ii) 
by modifying the gap width (along the $y$-direction) which changes 
the force integration interval. The spatial correlations decay 
rather slowly, meaning that tiny changes in the boundary conditions 
can not dramatically influence the pressure difference and, thus, 
the force. Still, there might be a visible impact on the effective 
force, since varying the particle elongation in the $y$-direction 
strongly changes the area on which the force is applied.

\end{document}